\newcommand{\cm}{\rm \,cm}
\newcommand{\mm}{\rm \,mm}
\newcommand{\m}{\rm \,m}

\newcommand{\s}{\rm \,s}

\newcommand{\sr}{\rm \,sr}

\newcommand{\MeV}{\rm \,MeV}
\newcommand{\eV}{\rm \,eV}

\newcommand{\GeV}{\rm \,GeV}

\newcommand{\anue}{$\bar{\nu}_e$ }
\newcommand{\numu}{$\nu_{\mu}$ }

\newcommand{\nutau}{$\nu_{\tau}$ }

\newcommand{\lsim}{\lower .5ex\hbox{$\buildrel < \over {\sim}$}}
\newcommand{\gsim}{\lower .5ex\hbox{$\buildrel > \over {\sim}$}}

\documentclass[11pt]{article}         
\usepackage{epsfig}                   
\begin{document}
\title{New MACRO results on atmospheric neutrino oscillations}
\author{ G. Giacomelli    and A. Margiotta \\
Dipartimento di Fisica  dell'Universit\`{a} di Bologna \\and INFN, Sezione di Bologna,  I-40127  Bologna, Italy\\
   giacomelli@bo.infn.it, margiotta@bo.infn.it \\  For
  the MACRO Collaboration\footnote{see Ref. [1] for a list of MACRO Authors
    and  Institutions}\\
\textit{   Invited Paper at the NANP03 Int. Conf., Dubna, 2003}}
\maketitle
\begin{abstract}
The final results of the MACRO experiment on atmospheric neutrino
oscillations are presented and discussed. 
The data concern different event topologies with average neutrino energies
of $\sim 3$ and $\sim 50$ GeV. Multiple Coulomb Scattering of the high
energy muons in absorbers was used to estimate the neutrino energy of each
event. The angular distributions, the $L/E_\nu$ distribution, the particle
ratios and the absolute fluxes all favour $\nu_\mu \rightarrow \nu_\tau$
oscillations with maximal mixing and $\Delta m^2 =0.0023 \: \rm eV^2$. A
discussion  is made on the Monte Carlos used for the atmospheric neutrino
flux.  Some results on neutrino astrophysics are also briefly discussed.\\
\it{Keywords: }\rm{ Astroparticle physics; underground detectors; neutrino oscillations.}
\end{abstract}
\section{Introduction}
MACRO was a large area multipurpose underground detector \cite{r1} designed
to search for rare events and rare phenomena in the penetrating cosmic
radiation.  The experiment  obtained important results on atmospheric
\numu{} oscillations and performed  \numu astronomy studies,  indirect
searches for WIMPs and  searches for low energy  \anue from stellar gravitational collapses \cite{r2}. 

The detector was located in Hall B of the underground Gran Sasso Lab at an average rock overburden of 3700 m.w.e.; it
started data taking with part of the apparatus in \( 1989 \); it was
completed in \( 1995 \) and  was running in its final configuration until
December  2000. 

\begin{figure}[th]
\vspace{1.cm}
  \centerline{\psfig{file=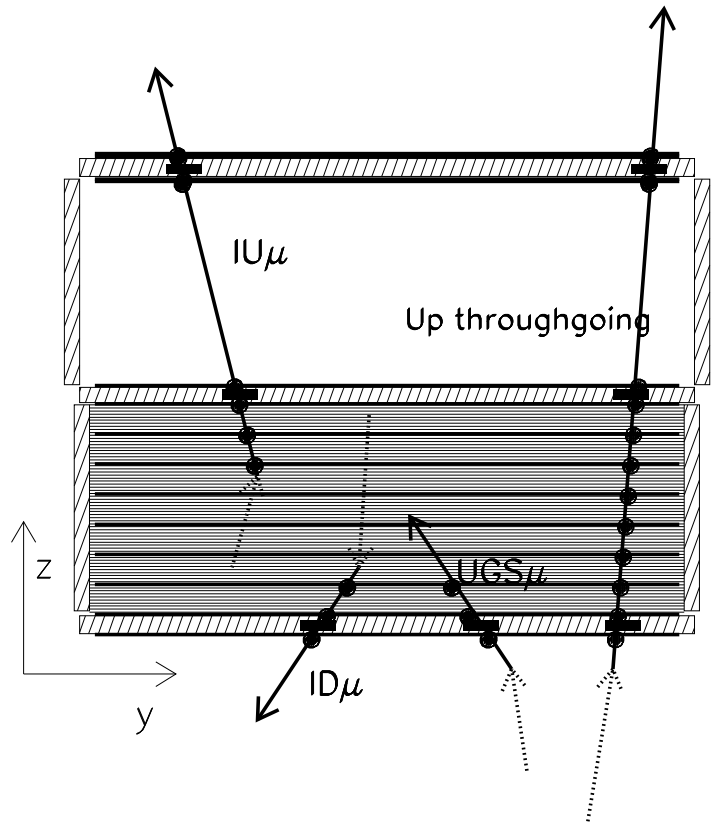,width=3.8in}}
\vspace*{8pt}
\caption{\label{fig1}\small  Cross section of the
  detector, which also shows the various atmospheric neutrino topologies
  measured. The horizontal scintillators are indicated as dashed boxes, the
  streamer tubes as horizontal thick black lines, the rock absorbers as
  boxes with horizontal thin lines. The sides of the detector were formed
  by 1  layer of scintillators and 6 layers of streamer tubes.}
\end{figure}

The detector had global dimensions of \( 76.6\times12 \times9 .3{\m }^{3}
\) and provided a total acceptance to an isotropic flux of particles of \(
\sim 10,000{\m }^{2}{\sr } \). The detector was composed of three horizontal layers of
liquid scintillation counters, 14 layers of limited streamer tubes and one layer of 
nuclear track detectors, Fig. \ref{fig1}. Vertically it was divided into two parts: the lower part
contained 10 layers of streamer tubes, 7 layers of rock absorbers and 2
layers of liquid  scintillators;  the upper part was empty, contained the
electronics and was  covered by 1 layer of scintillators and 4 layers of
streamer tubes. The sides of the  detector were covered with 1 layer of
scintillators and 6 layers  of limited streamer tubes, so as to obtain a closed box structure.
Each of the subdetectors could be used in \lq\lq stand-alone\rq\rq~and in \lq\lq combined\rq\rq~mode.

In the following we shall briefly recall neutrino oscillation formulae and
atmospheric neutrinos; then we shall discuss our main results including the
ratio of vertical to horizontal muons and the study of the $L/E_\nu$ distribution.
After a detailed analysis of the present situation of the different Monte
Carlos  (MCs), we summarize our atmospheric neutrino oscillation results
and we conclude with a short discussion of some neutrino astrophysics studies.
\section{Atmospheric neutrino oscillations. Monte Carlos}
\textbf{Neutrino oscillations.} If neutrinos have non-zero masses, one has
to consider the 3 weak \textit{flavour eigenstates} $\nu_e$, $\nu_\mu$,
$\nu_\tau$  and the 3 \textit{mass eigenstates} $\nu_1$, $\nu_2$,
$\nu_3$. The  flavour eigenstates $\nu_\textit{l}$ are linear combinations
of the mass  eigenstates $\nu_\textit{m}$. Neutrino oscillations depend on
six independent  parameters: two mass-squared differences, $\Delta
m^{2}_{12}$ and  $\Delta  m^{2}_{23}$, three mixing angles $\theta_{12}$,
$\theta_{13}$,  $\theta_{23}$ and the CP-violating phase $\delta$. In the
simple case of  two flavour eigenstates ($\nu_\mu$, $\nu_\tau$) which
oscillate with two mass  eigenstates ($\nu_2$, $\nu_3$), and $\delta = 0$ one has:
\begin{equation}
         \nu_\mu = \nu_2 \cos \theta_{23} + \nu_3 \sin \theta_{23} 
\end{equation}
\begin{equation}
        \nu_\tau = -\nu_2 \sin \theta_{23} + \nu_3 \cos \theta_{23}
\end{equation}
The survival probability of a $\nu_\mu$ beam is 
\begin{equation}         
P(\nu_\mu \rightarrow \nu_\mu )
 \simeq 1 - \sin^{2} 2 \theta_{23} \sin^{2} \left (\frac{1.27 \Delta m^{2}_{23}\cdot L}{E_{\nu}} \right)
\label{eq:2}
\end{equation}
where $\Delta  m^{2}_{23} = m^{2}_{3} - m^{2}_{2}$ and L is the distance
travelled  by the neutrino of energy $E_{\nu}$ from production to
detection. The two-neutrino approximation is adequate for discussing our data.

\textbf{Atmospheric neutrinos.}
High energy (HE) primary Cosmic Rays (CRs), protons and nuclei, interact in
the upper atmosphere producing a large number of pions and kaons, which
decay yielding muons and muon neutrinos; then muons decay yielding
$\nu_\mu$'s and $\nu_e$'s. The ratios of their numbers are
$N_{\nu_{\mu}}/N_{\nu_{e}}\simeq 2$ and $N_{\nu}/N_{\overline{\nu}}\simeq
1$. The atmospheric neutrinos are produced in a spherical shell at
about 10-20 km above ground and proceed towards the Earth.
Atmospheric neutrinos have energies from a fraction of GeV up to more than
100 GeV and they travel distances L from few tens of km up to $\sim 13000$
km. An underground detector is ''illuminated'' by a flux of neutrinos from
all  directions and it can make oscillation studies for $1 < L/E_\nu< 10^4 \: \rm km/GeV$.

MACRO detected upgoing muon neutrinos via charged current interactions, \(
\nu _{\mu } \rightarrow \mu \); the upgoing muons were identified with the
streamer tube system (for tracking) and the scintillator system (for
time-of-flight measurement).
The events measured and expected for the three measured topologies,
indicated in Fig. \ref{fig1} and the $L/E_\nu$ distribution deviate from
Monte  Carlo expectations without oscillations,
Figs.\ref{fig2}, \ref{fig4} and Tables 1, 2; the deviations point to the same ${\nu_\mu \rightarrow
  \nu_\tau}$  oscillation scenario \cite{r2}-\cite{r8}.

\textbf{Monte Carlos.} The measured data of Fig. \ref{fig2} were compared
with  different MC simulations.
In the past  we used the neutrino flux computed by
the Bartol96 group \cite {bartol} and the GRV94  \cite {gluck} parton
distribution. For the low energy channels the cross sections in
\cite {cross_sec} were used; the propagation of muons to the detector was
done using the energy loss calculation by Lohmann et al. \cite
{lohmann}. The total systematic uncertainty in the predicted flux of
upthroughgoing muons, adding in quadrature the errors, was estimated to be
$\pm 17 \:\%$;  this is mainly a scale error that does not change the shape
of the angular distribution. The error on the shape of the distribution is
$ \sim 6 \%$. The detector was simulated using GEANT \cite {geant}. A
similar MC (Honda96)  was used by the Superkamiokande Collaboration \cite {honda}.

Recently new improved MC predictions were made available by the Honda \cite {honda} and Fluka
\cite {fluka} groups. 
They include three dimensional calculations of hadron production and decays
and of neutrino interactions, improved hadronic model and
new fits of the primary cosmic ray flux. The two MCs yield predictions for
the non oscillated and oscillated \numu fluxes in complete agreement with
each other, equal to within few \%, see Figs. \ref{fig3}a, b. The shapes of the
angular distributions for oscillated and non oscillated Bartol96, new Fluka
and new Honda fluxes are the same to within few \%, Fig. \ref{fig3}a, b.
The absolute values of our upthroughgoing muon data are about $20 - 30 \%$
above those  predicted by the new Fluka and Honda MCs; this situation is
also true for the  new Superkamiokande data \cite{hayato}; the measurements
of the flux value at $\cos \vartheta = 0$  performed by various
experiments do not seem to be conclusive \cite {cos_0}. Note in
Fig. \ref{fig3}a that   the new Fluka muon flux
with an older cosmic ray fit is considerably above the  new Fluka flux with the new CR fit.
Our high energy \numu data suggest that the new Honda and Fluka predictions should be raised
by about $20 - 25 \%$,  probably because of the used new CR fit.
The inclusion of the new ATTIC Collaboration measurements of primary CRs
seems to lead to the old energy dependence of $E^{-2.71}$ \cite{battiston}.
Thus the Bartol96 MC may  probably still be used for the prediction of the
absolute flux, besides predicting well the shape of the angular
distribution. In the following we shall  use mainly the predictions of this MC.
It should be noted that the evidence for neutrino oscillations rests
primarely with the shape  of the angular distribution and this is the same
in all MC calculations.

\section{MACRO results on atmospheric neutrinos.} 
\begin{figure}[th] \centerline{\psfig{file=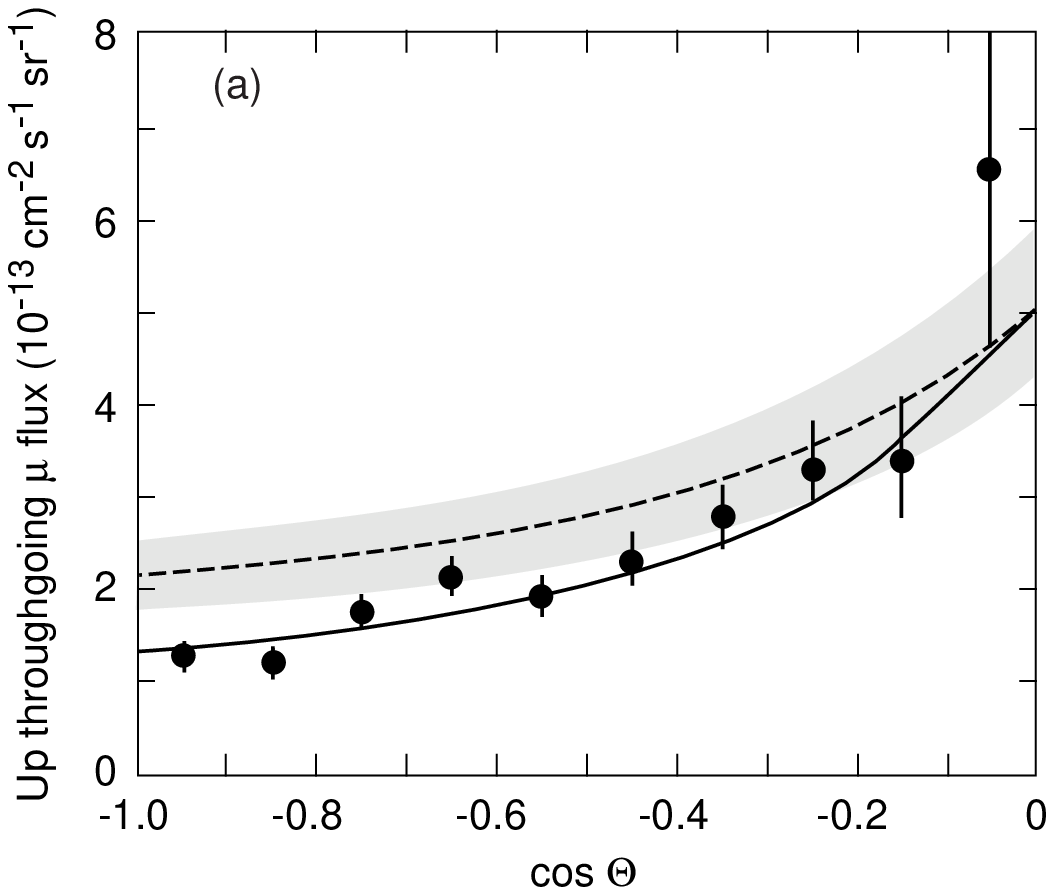,width=2.4in}
    \hspace{.5cm} \psfig{file=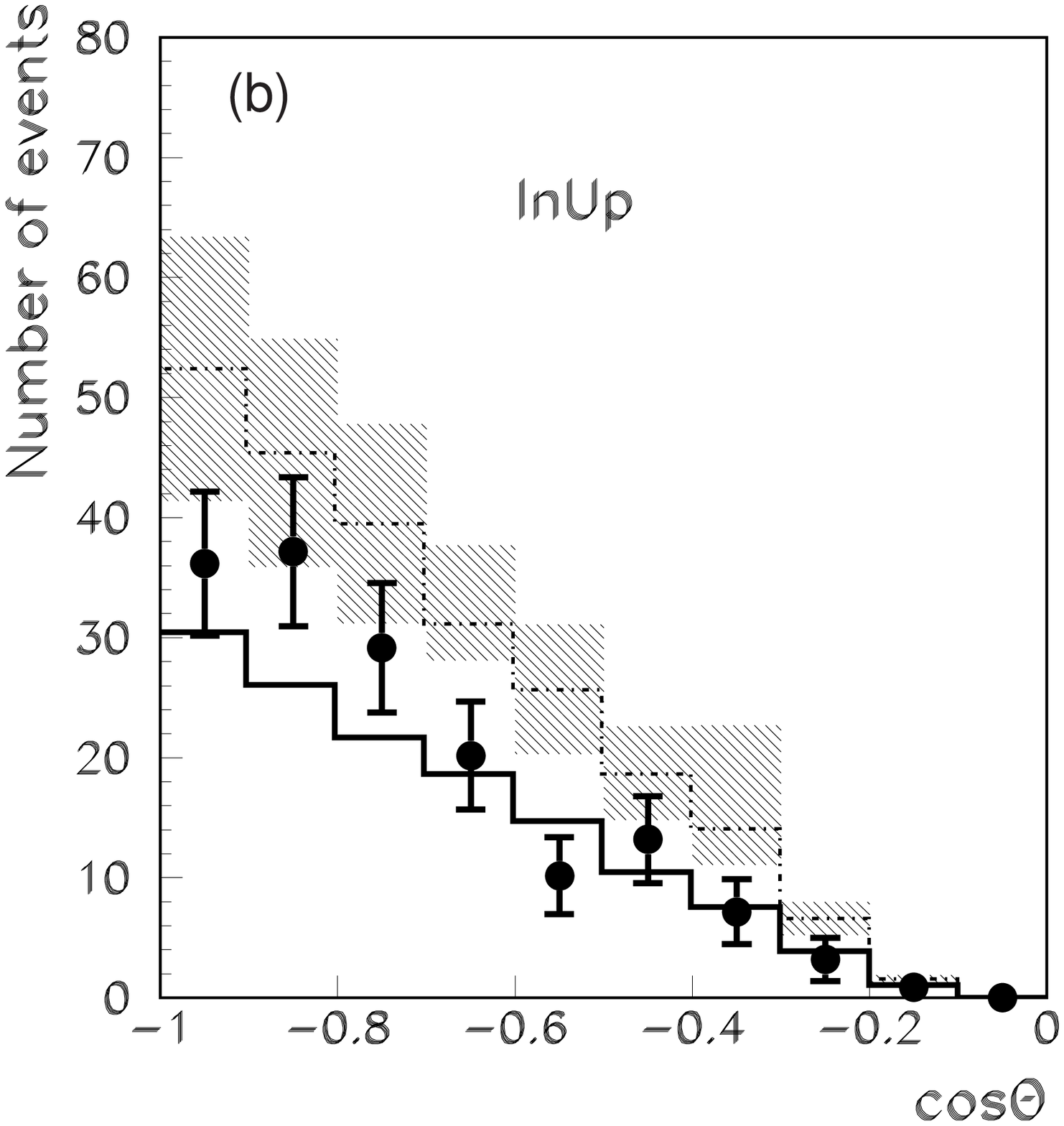,width=2.4in}
    \psfig{file=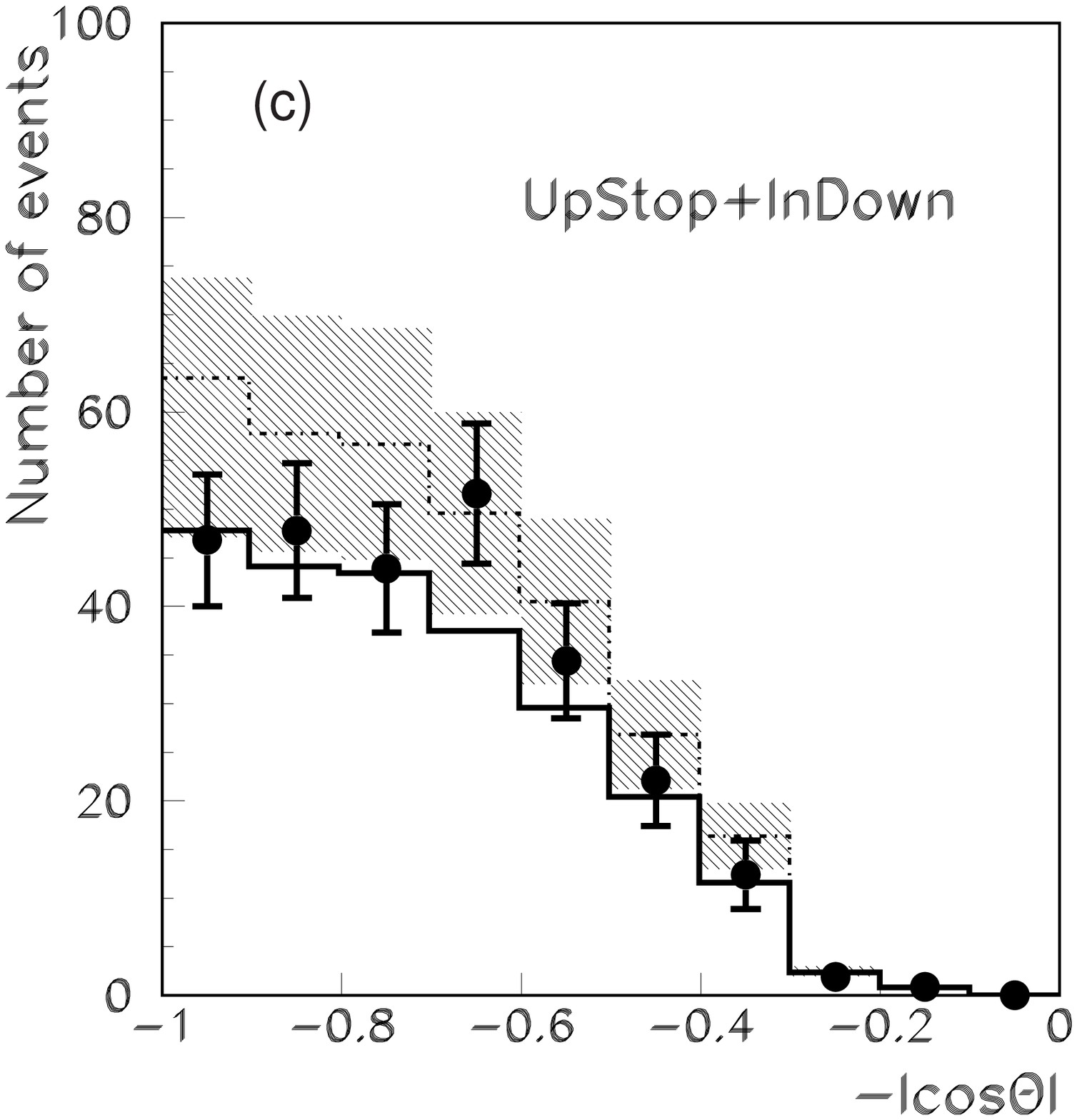,width=2.4in}}
\caption{\label{fig2}\small Zenith distributions for the MACRO data (black
  points) for (a) upthroughgoing, (b) semicontained and (c) up-stopping
  muons +  down semicontained. The dashed lines are the no-oscillation MC
  predictions Bartol96 in (a) and new FLUKA in (b) and (c) (with scale error bands); the solid lines refer to 
 \( \nu _{\mu }\rightarrow  \nu _{\tau } \) oscillations  with maximal
  mixing and $\Delta  m^{2} = 2.3 \cdot 10^{-3}$ eV$^{2}$.}
\end{figure}

\begin{figure}[th]
  \centerline{\psfig{file=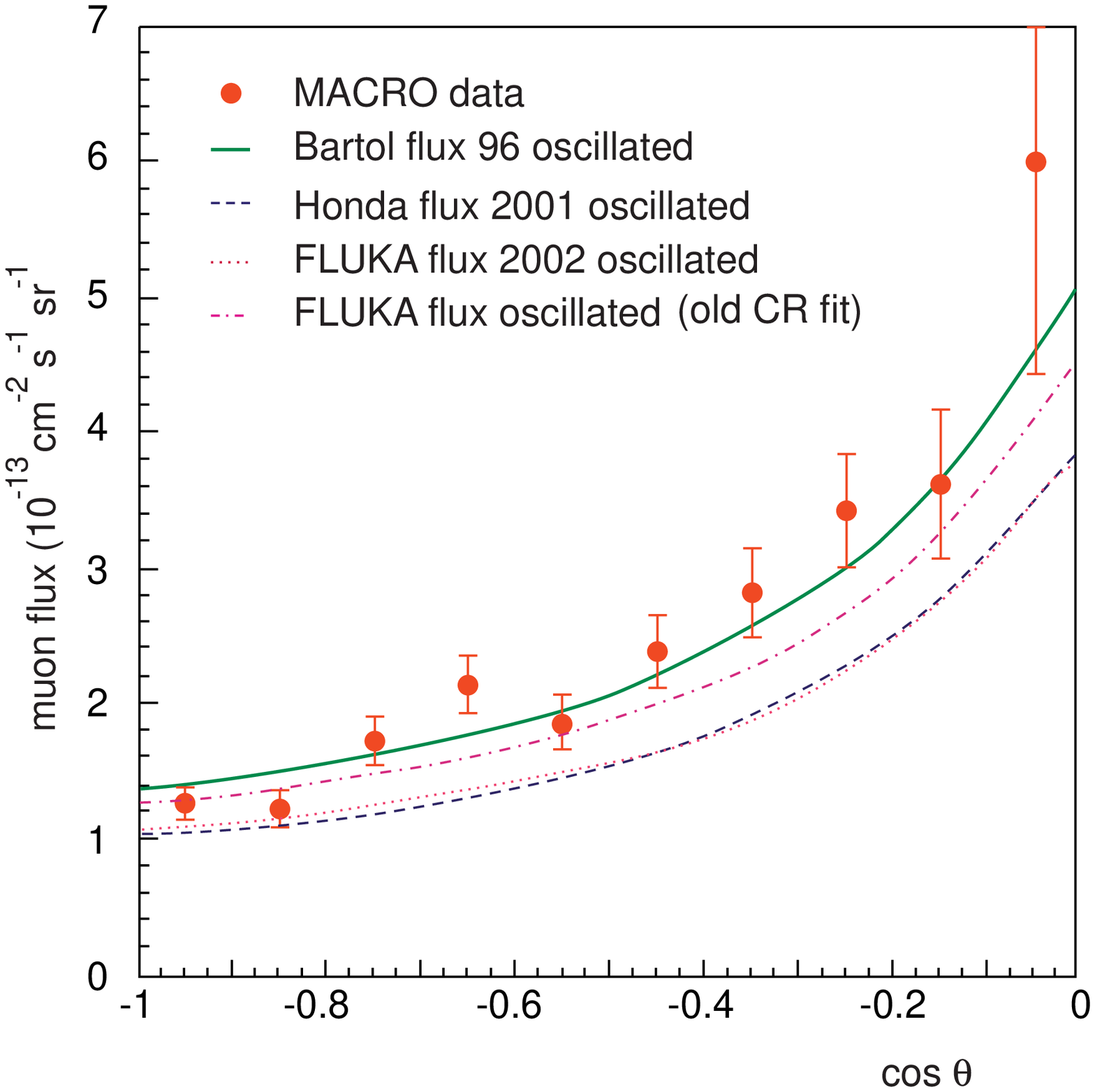,width=3.2in}\psfig{file=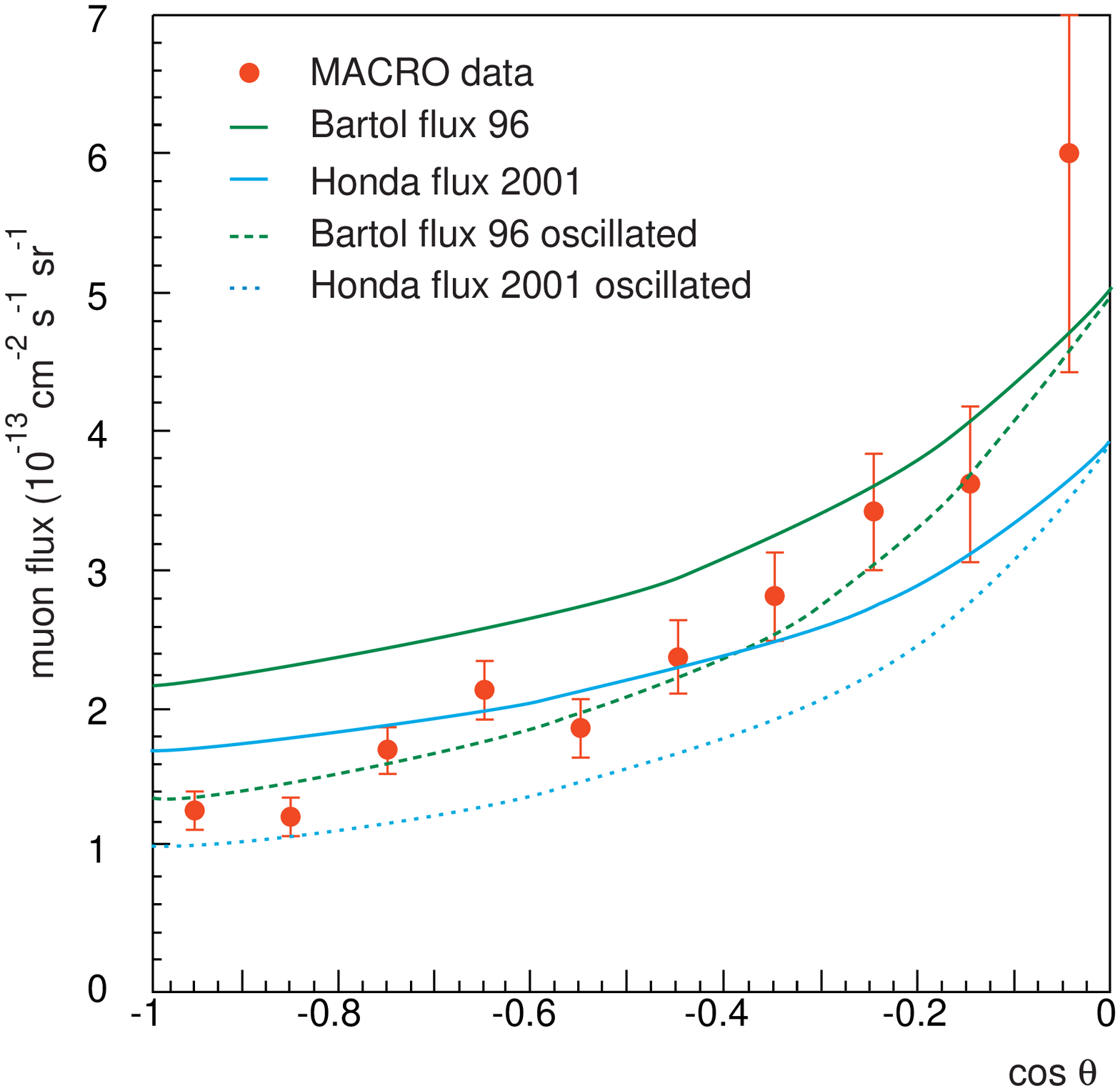,width=3.2in}}
 \hspace{3.7cm}(a)  \hspace{7.7cm} (b) 
\caption{\label{fig3}\small (a) Comparison of the measured angular
  distribution for  upthroughgoing muons and the oscillated flux obtained
  from different MCs (for  $\Delta  m^{2} = 2.3 \cdot 10^{-3}$ eV$^{2}$ and
  maximal mixing). The solid line is the  Bartol96 flux, the dotted and the dashed
  lines at the bottom  are the new Fluka and Honda fluxes; the dotted-dashed line
  in the middle is the Fluka  flux using a different (older) fit to the
  cosmic ray data.  (b) Comparison of our measurements with  the Bartol96
  and the new Honda 2001 oscillated  and non oscillated fluxes.}
\end{figure}

The {\it upthroughgoing muons } come from $ \nu _{\mu } $ interactions
in the rock below the detector; the \( \nu _{\mu } \)'s have a median
energy \( \overline{E}_{\nu }\sim \, \, 50\, \, \GeV  \); muons with \(
E_{\mu }>1\GeV  \) cross the whole detector. The data taking livetime was
slightly over 6 years (full detector equivalent). The data, Table 1 and 
Fig. \ref{fig2}a, deviate in shape and in absolute value from the Bartol96 MC
predictions (see  Section 2). This was first pointed by MACRO in 1995 \cite{r4}.

\begin{figure}[th]
   \centerline{\psfig{file=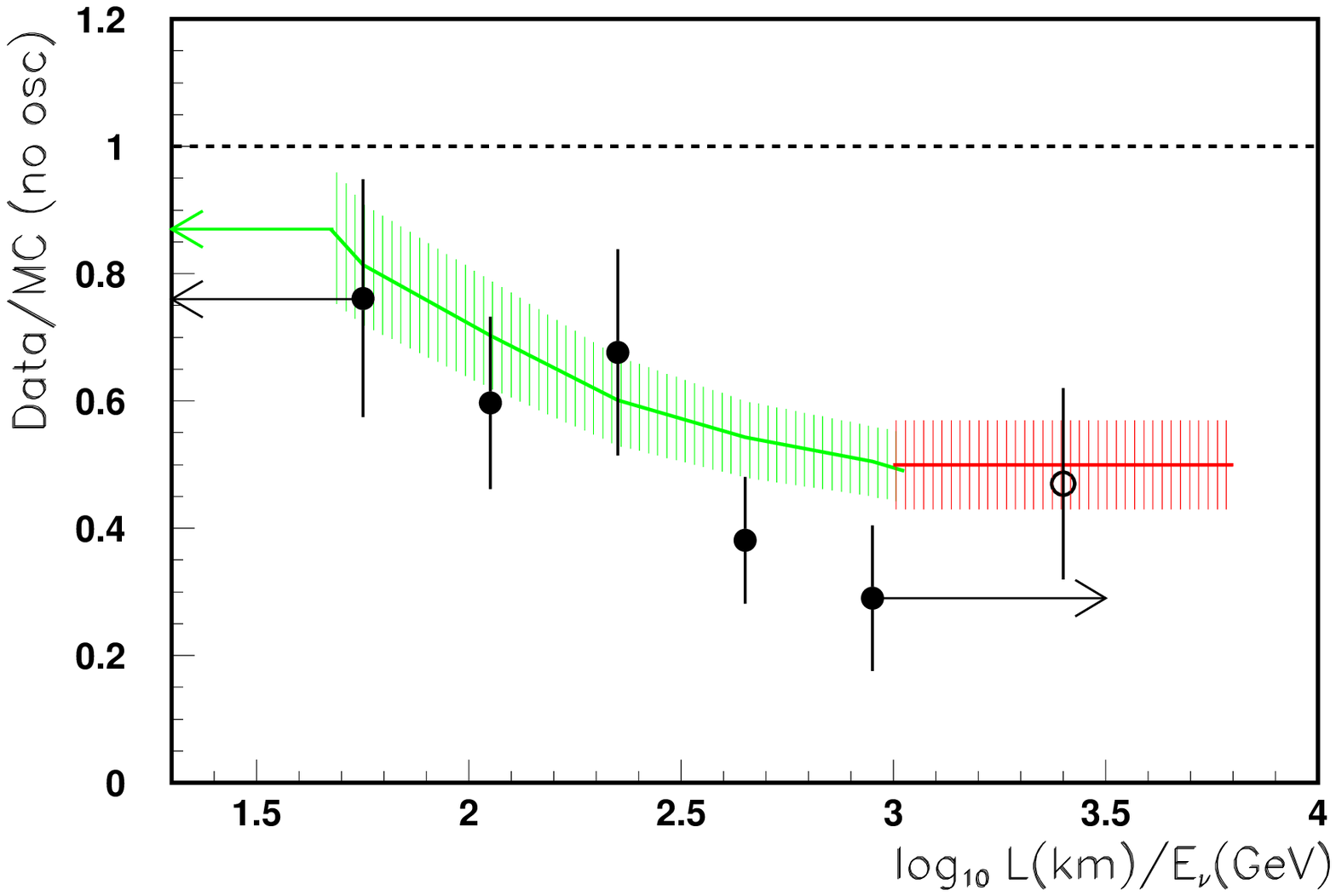,width=4.5in}}
\vspace*{8pt}
 \caption{\label{fig4}\small
Ratio (Data/MC Bartol96) as a function of estimated $L/E_{\nu}$ for the upthrougoing
muon sample  (black circles) and the semicontained up-$\mu$ (open
circle). For  upthroughgoing muons the muon energy was estimated by
multiple Coulomb scattering  and $E_{\nu}$ by MC methods. The shaded
regions represent  the uncertainties in the MC predictions assuming $\sin^2
2 \theta=1$ and $\Delta m^2=0.0023$ eV$^2$. The  horizontal dashed line at
Data/MC=1 is the  expectation for no oscillations. }
\end{figure}
A large number of possible systematic effects and backgrounds
that could affect the measurements were studied. One of the most significant checks was
performed using only the scintillator system with two different and
independent electronic systems \cite{r3,r8}. Two different final analyses
of the upthrough data have been performed: they agree to within $5\%$.

Assuming no oscillations, the number of expected upthroughgoing muon events integrated over all
zenith angles from Bartol96 without oscillations is 1169;  the measured
number is 857, Table 1. Thus the ratio of the
observed number of events to the Bartol96 expectation is  0.73.

\textbf{\numu $\rightarrow$ \nutau versus \numu $\rightarrow \nu_s$}. 
Matter effects due to the difference between the weak interaction effective
potential for muon neutrinos with respect to sterile neutrinos ($\nu_s$)
would produce a different total number and a different zenith angle distribution
of upthroughgoing muons.
The ratio $R_1 = $ Vertical/Horizontal $= N(-1<cos\theta < -0.7) / N(-0.4 <
cos\theta < 0)$ was  used to test the \numu $\rightarrow
\nu_s$ oscillation hypothesis versus \numu $\rightarrow$ \nutau \cite{r2}, \cite{r6}, \cite{r8}. 
The measured value is $R_{meas} = 1.38$; it should be compared
to $R^{min}_{\tau} = 1.61$ and $R^{min}_{sterile}= 2.03$, which are the
minimum values of the ratios for \numu $\rightarrow$ \nutau and \numu
$\rightarrow \nu_{sterile}$ oscillations for $ \Delta
m^{2}_{23}=2.3\cdot 10^{-3}\, \, \eV ^{2} $ and maximal mixing.
\begin{figure}[th]
  \centerline{\psfig{file=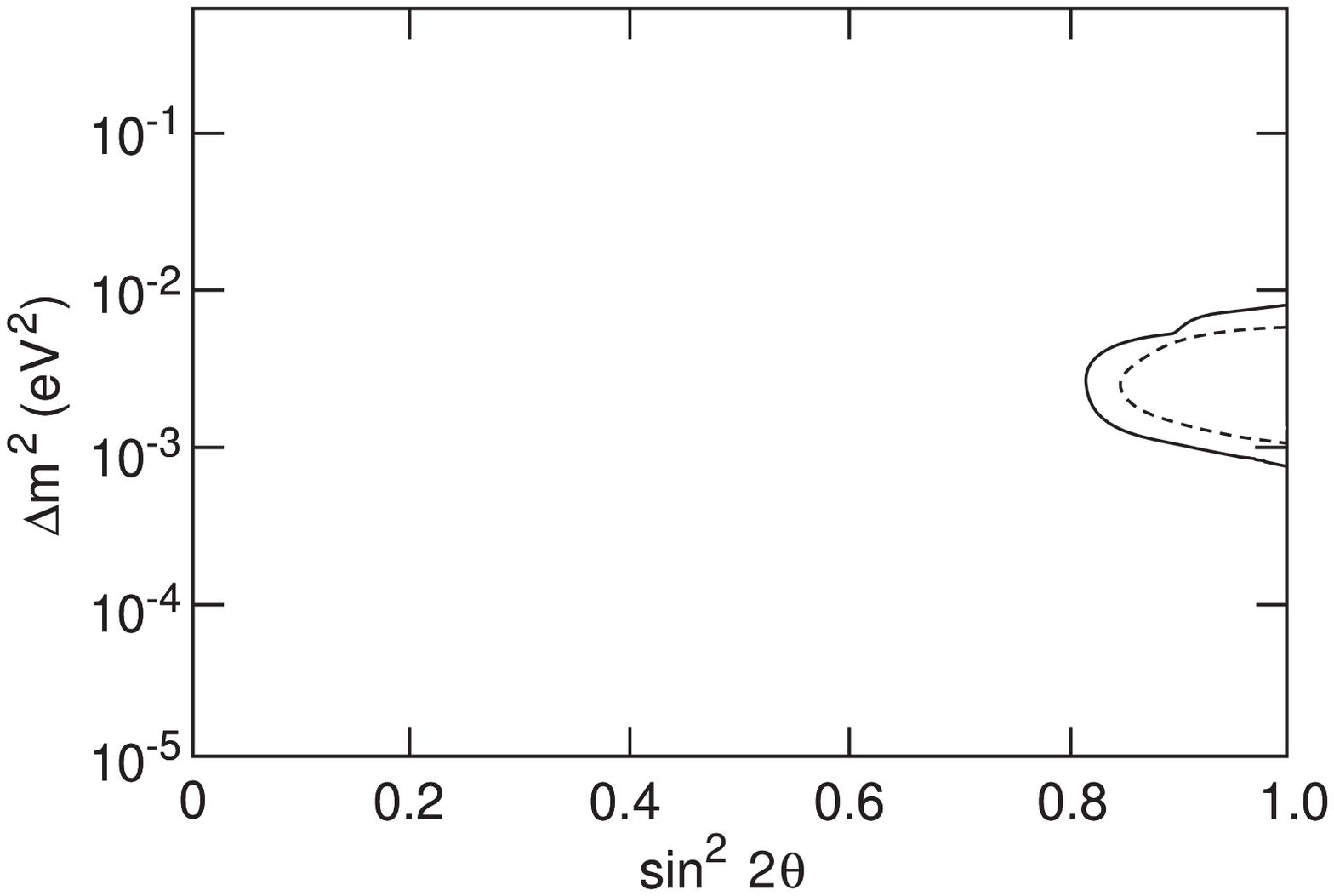,width=3.2in}\psfig{file=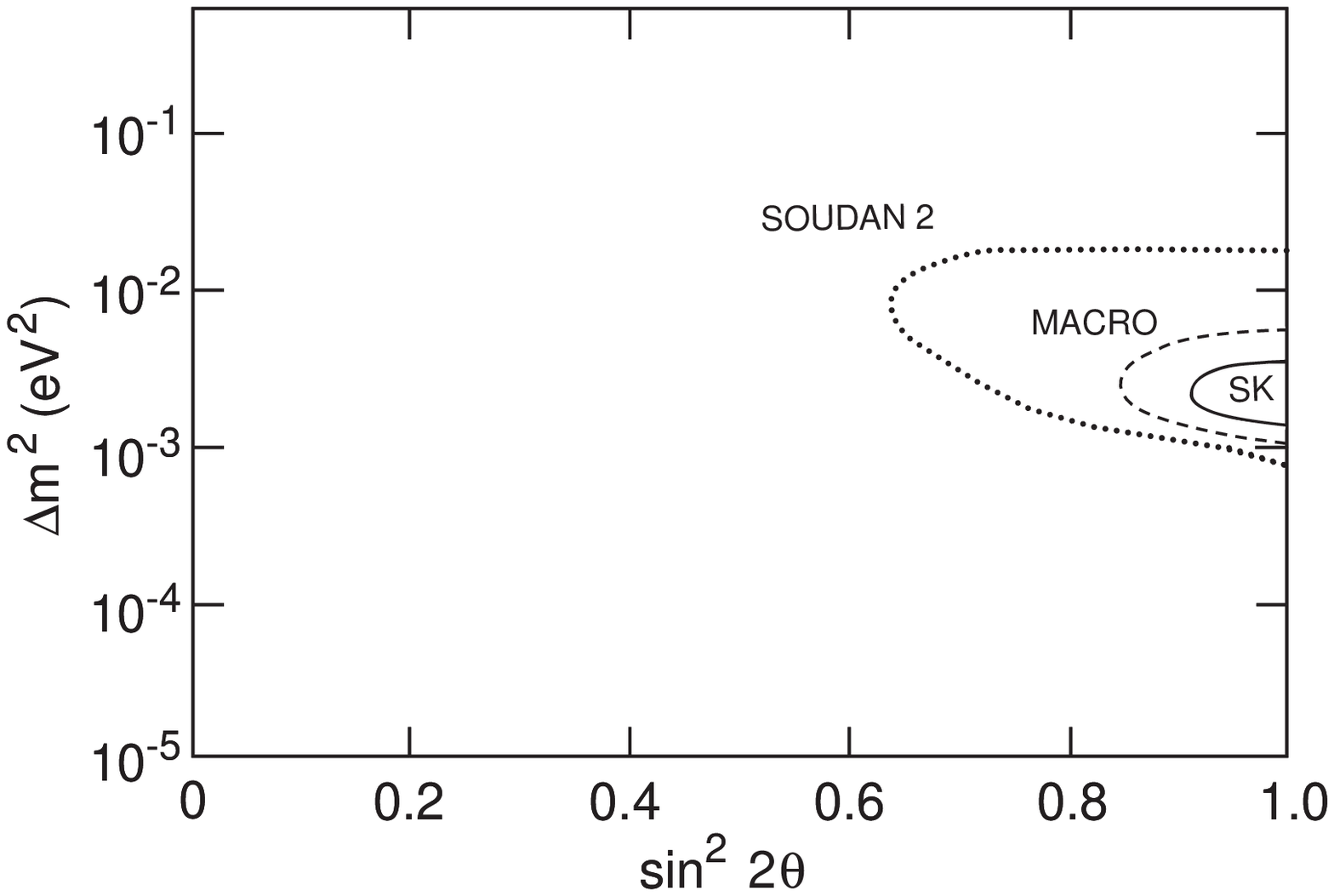,width=3.2in}}
   \hspace{3.3cm}(a)  \hspace{9.2cm} (b) 
\vspace*{8pt}
\caption{\label{fig5}\small  
Interpolated qualitative 
90\% C.L. contour plots of the allowed regions in the \( \Delta m^{2} - \sin^{2}2\theta \) plane (a) for the MACRO data using only the ratios $R_{1}, R_{2}, R_{3}$ (outer continuous line) and using also the absolute values assuming the validity of the Bartol96 fluxes. (b) Comparison of the allowed regions for SK, MACRO and Soudan 2 (All contours have been smoothed out and must be considered as qualitative).
}
\end{figure}
\begin{table}
\vspace*{8pt}
{\centering \begin{tabular}{ccc}
\hline 
Topology                          & Measured events & $ (Data/MC)_{no osc}$ \\ \hline
Upthroughgoing muons              & 857 & 0.73    \\ 
Semicontained (UP)                & 157 & 0.55     \\
Upstopping + Down                 & 262 & 0.74    \\ 
semicontained (ID + UGS)          &     &       \\ \hline
\end{tabular}\par}
\caption{\small For each topology the table gives the number of events 
  measured and the ratios $(Data/MC)_{no osc} $ using the
  Bartol96 MC.}
\end{table}
\begin{table}
{\centering \begin{tabular}{cccc}
\hline 
Ratio                  & $R_{meas} \pm \sigma_{stat}$ & $R_0$  & $R_0 \le R_{meas}$ probab\\
                       &                              &        & \it{a priori} II         \\ \hline
Vertical/Horizontal    & $1.38 \pm 0.12$              & $2.11$ & $6.4 \times 10^{-6} $    \\ 
$N_{low}/N_{high}$     & $0.85 \pm 0.16$              & $1.5$  & $7.7 \times 10^{-3} $    \\
$InUp/(InDown+UpStop)$ & $0.60 \pm 0.06$              & $0.745$& $3.1 \times 10^{-2} $    \\ \hline
Combination            &                              &        & $3.4 \times 10^{-7} $    \\ \hline               
\end{tabular}\par}
\caption{\small Ratios between different event categories: $R_{meas}$ is
  the measured value, $R_{0}$ is  the expected value from the Bartol96 MC
  without oscillations; using the new FLUKA MC the ratios differ by at most $5 \%$.  The last column gives the one-sided probability
  compatible with a  statistical fluctuation computed according to the
  \textit{a priori} assumptions  explained in the text.
}
\end{table}
For \numu $\rightarrow $ \nutau, the one-sided probability to measure a
value lower than  $R_{meas} $ is $7.2\%$.
For \numu $\rightarrow \nu_s$ the probability is $P^{best}_{ster} \simeq
1.5 \cdot 10^{-4}$; $P^{best}_{\tau}$ is about 480 times larger than
$P^{best}_{ster}$. Therefore  \numu $\rightarrow \nu_s$ oscillations (with
any mixing) are excluded at about  $99.8\%$ c.l. with respect to \numu
$\rightarrow$ \nutau oscillations with  maximal mixing \cite{r8}.

\textbf{Oscillation probability as a function of the ratio
  $L/E_\nu$}. $E_\nu$ was estimated by measuring the muon energy, $E_\mu$, by means of 
 the muon Multiple Coulomb Scattering (MCS) in the rock absorbers in
  the lower MACRO.  The best  method employed the
  streamer tubes in {}``drift mode{}'', using the  special electronics designed
  to search for magnetic monopoles \cite{mono}. The space resolution achieved was \(
  \simeq 3{\mm } \). For each muon, seven variables were given in input to
  a Neural Network (NN) previously trained to estimate muon energies with
  MC events of known input energy crossing the detector at different zenith
  angles. The distribution of the ratio $R = (Data/MC_{no osc})$ obtained
  by this analysis is plotted in Fig. \ref{fig4} as a function of $
  (L/E_\nu)$ \cite{r7}. Notice that the data extend from $(L/E_\nu) \sim
  30$ km/GeV to $ 5000$ km/GeV.

The \textit{Internal Upgoing} (IU) muons ($\sim $ 157 events, 285 expected by Bartol96 and 235 expected by new FLUKA,
Table 1)
come from  $\sim 4$ GeV $ \nu _{\mu }$'s interacting in the lower
apparatus,  Fig. \ref{fig1}. Compared to the no-oscillation prediction there is a reduction of about a factor of two in
the flux of these events, without any appreciable distortion in the shape of the zenith
distribution, Fig. \ref{fig2}b \cite{r5}, \cite{r8}. The MC predictions for
no oscillation in Figs. \ref{fig2}b  and \ref{fig2}c are given by the
dashed lines with a $21 \:\%$  systematic scale band. Notice that at these
energies, the new Honda and Fluka MC predictions are low by about $12 \%$.

The \textit{upstopping muons} (UGS) are due to $\sim 3$ GeV  \( \nu _{\mu } \)'s
interacting below the detector and yielding upgoing muons stopping in the
detector. The \textit{semicontained downgoing muons} (ID) are due to \( \nu
_{\mu } \)-induced downgoing muon tracks with vertex in the lower MACRO. 
The two types of events (262 compared to 354 expected by Bartol96 and 315 by FLUKA, Table 1) are identified by means of
topological criteria; the lack of time information prevents to distinguish
the two sub-samples.  The upgoing $\nu_{\mu}$'s should all have oscillated
completely, while  the downgoing \numu do not.
The zenith distribution shows, as expected, a uniform deficit of about 25
\%  of the measured number of events
with respect to the no-oscillation prediction, Fig. \ref{fig2}c \cite{r5}, \cite{r8}. 

\section{Determination of the oscillation parameters}
In the past, in order to determine the oscillation parameters, MACRO made fits
to the shape of the upthroughgoing muon distribution and to their absolute
flux compared to the Bartol96 MC prediction (with a $ 17 \%$ systematic
scale uncertainty). The other data were only used to verify the consistency
and make checks. The result  was $\Delta m^{2} = 0.0025$ eV$^2$ and maximal mixing. 

Later, in \cite{r6,r7} the high energy data for the zenith
distribution, $R_{1} =(N_{vert}/N_{horiz})$, were combined with the HE neutrino energy 
measurement, $R_{2}=N_{low}/N_{high}$. They are independent of the
neutrino absolute flux; with these 
two informations the significance of observation of neutrino
oscillations is about 4.7 $\sigma$. 

In order to reduce the effects of possible uncertainties in the MCs (to
about $6\%$) we now use  the following three independent ratios \cite{r8}.
\begin{enumerate}
        \item [(i)] High Energy Data: zenith distribution ratio: $R_{1} = N_{vert}/N_{hor}$
        \item [(ii)] High Energy Data, neutrino energy measurement ratio: $R_{2} = N_{low}/N_{high}$
        \item [(iii)] Low Energy Data: Ratio $R_{3} = (Data/MC)_{IU}/(Data/MC)_{ID+UGS} $.
\end{enumerate}
 
Table 2 gives for each ratio the measured value, $R_{meas} $, the
prediction, $R_0 $, for no oscillations using either the Bartol96 or the new FLUKA MC; the
probability value for no oscillations  obtained allowing the predicted
ratio to fluctuate around the mean  value (considering the total number of
events equal to the  predicted value without oscillations). The no
oscillations hypothesis has a probability  P  $ \sim 3.4 \times 10^{-7}$
(Table 2),  ruling out the no-oscillation hypothesis by  $ \sim 5 \sigma$.
The formula used for  combining independent probabilities is $P =
P_{1}P_{2}P_{3} (1  - \ln P_{1}P_{2}P_{3} + 1/2 (\ln P_{1}P_{2}P_{3})^2)$
\cite{roe}.

By fitting the three ratios  to the \numu $\rightarrow$ \nutau oscillation
formulae we  obtain $\sin^{2} 2\vartheta = 1$,   $\Delta  m^{2} = 2.3
\cdot 10^{-3}$ eV$^{2}$ and the allowed region indicated in Fig. \ref{fig5}a. 

We could also add the information on the absolute flux values of the
\begin{enumerate}
   \item[(iv)] high energy data, assuming the validity of the Bartol96 flux
     and a systematic scale error of $17 \%$, $R_{4} =
     N_{meas}/N_{Bartol-nooscill} \simeq 0.73$, $P_{4}\simeq 0.07$
   \item [(v)] low energy semicontained muons, with a systematic scale
     error of $21 \%$, \\
     $R_{5}= N_{meas}/N_{Bartol-nooscill} \simeq 0.67$, $P_{5} \simeq 0.08$
     (At these low energies the Bartol96, Honda  2001 and Fluka 2002 MC
     neutrino fluxes are  essentially equal).
\end{enumerate}

Fig. \ref{fig5}a shows the MACRO allowed regions for the \( \nu _{\mu
  }\rightarrow \nu _{\tau } \) oscillation parameters in the \(\Delta m^{2} -
\sin^{2}2\theta \) plane using only the three ratios $R_1$,  $R_2$ and
  $R_3$ (continuous line) and using also the two estimated absolute values,
  assuming the validity of the Bartol96  fluxes (dashed line). In
  Fig. \ref{fig5}b are compared the MACRO,  SuperKamiokande (SK) \cite{hayato} 
and Soudan 2  \cite{allison} allowed regions. All limiting lines are
  qualitative  smoothed interpolations.

\section{Neutrino Astrophysics with MACRO}
\textbf{Search for Astrophysical Sources of High Energy Muon Neutrinos.} 
High energy \( \nu _{\mu } \)'s are expected to come from several
galactic and extragalactic sources. Neutrino production requires
astrophysical accelerators of protons and astrophysical "beam
dumps". 
A sensitive search was made for upgoing muons produced by neutrinos coming from
celestial sources, interacting below the detector.
90\% c.l. upper
limits were established on the muon fluxes from
specific celestial sources; the limits are in the range \(
10^{-15}-10^{-14}{\cm }^{-2}{\s }^{-1} \).
In the case of GX339-4 ($\alpha = 255.71^o $, $\delta= -48.79^o $)
and Cir X-1  ($\alpha = 230.17^o $, $\delta= -57.17^o $), we have  7 events:
they have been considered as background, therefore the upper flux limits
are higher; but the events could also be indication of signals
\cite{r10,r2,r3}. The pointing  capability of the detector was tested by
determining the shadows  of the Moon and of the Sun on the primary cosmic ray flux \cite{r29}

A search for time coincidences of the upgoing muons with \( \gamma  \)-ray
bursts was made. No statistically significant time correlation was found.
A search was also made for a diffuse astrophysical neutrino flux for
which an upper limit at the level of \( 1.5\cdot
10^{-14}{\cm }^{-2}{\s }^{-1} \) was established \cite{r11}.

\textbf{Indirect Searches for WIMPs.}Weakly Interacting Massive Particles (WIMPs) could be part of the
galactic dark matter; they could be intercepted by celestial bodies,
slowed down and trapped in their centers, where WIMPs and anti-WIMPs could
annihilate and yield neutrinos of multi\GeV{} energy,
in small angular windows from their centers. One of the best WIMP candidate could be the lowest mass neutralino.

A search was made for \numu $\rightarrow \mu $ coming from the center of
the Earth using $10^{o}- 15^{o}$ cones around the vertical; we obtained  muon flux limits of
\( \sim 10^{-14}{\cm }^{-2}{\s }^{-1} \).
The limits were compared with the predictions of a supersymmetric model by Bottino et al.;
our data eliminate a sizable range of parameters used in the model
\cite{r12,r3}. A similar procedure  was used to search for muon neutrinos from the
Sun: the upper limits are at the level of about \( 1.5\cdot 10^{-14}{\cm }^{-2}{\s }^{-1} \) \cite{r12}.

\textbf{Neutrinos from Stellar Gravitational Collapses.} A stellar
gravitational collapse of  the core of a massive star is expected to yield
a  burst of all types of neutrinos  and antineutrinos with energies of \(
5-30 \)~\MeV{} in \( \sim 10{\s } \). The \anue's  can be detected via \(
\bar{\nu }_{e}+p\rightarrow n+e^{+} \)  in the liquid scintillator.
About 120 \anue{} events could have been detected in our
580 t liquid scintillator for a stellar collapse at the center of our Galaxy.
Two separate electronic systems were used. 
Both had an energy threshold of \( \sim 7{\MeV } \) and recorded
pulse shape, charge and timing informations of the positron. Following  a \( >7{\MeV } \)
trigger, the first  system PHRASE lowered its threshold to $ \sim 1 {\MeV }$ for
\( 800 \, {\mu \s } \) in order to detect (with a \( \simeq 25\, \% \)
efficiency) the \( 2.2{\MeV } \) \( \gamma  \) released in the capture reaction
\( n\, +\, p\rightarrow d\, +\, \gamma  \) induced by
the neutron produced in the primary process.
A redundant supernova alarm system was in operation, alerting the
physicists on shift; a  procedure was  defined to alert the various supernova
observatories around the world \cite{r20}.
MACRO was completed at the end of 1994; the first parts of the detector started operation
in 1989. No stellar gravitational collapses in our Galaxy were observed from \( 1989 \) to  2000.

\section{Conclusions}
The MACRO detector took data from 1989 till the end of year 2000.
Important results were obtained in all the items  listed in the proposal. 

In particular for atmospheric neutrino oscillations we analyzed different
  event topologies,   different energies, exploited muon Coulomb multiple
  scattering in  the  detector to  measure muon energies and studied the
  effects of using different MCs. All data are in agreement  with the hypothesis of $\nu_\mu \rightarrow
  \nu_\tau$ oscillations, with maximal mixing and $\Delta m^{2}_{23} \simeq 2.3
  \cdot 10^{-3}$ eV$^{2}$, and rule out the no oscillation hypothesis by  $ \sim 5 \sigma$.
 
Studies were made on muon neutrino astronomy \cite{r10}, \cite{r29}, seasonal, solar and
  sidereal variations \cite{r2} \cite{side}, and on  possible Dark Matter
  candidates :  WIMPs (looking for muon neutrinos from the centers of the Earth and of the
  Sun),  Nuclearites and  Q-balls  \cite{r2}, \cite{r18}.   
In the search for GUT Magnetic Monopoles MACRO obtained the best existing direct flux
upper limit ($\Phi <  1.4\times 10^{-16} cm^{-2} s^{-1}sr^{-1}$) over the
widest $\beta$ range  ($4 \times 10^{-5} < \beta < 1$) \cite{mono}.
\section*{Acknowledgments}
We would like to acknowledge the cooperation of all the members of the
MACRO collaboration. MACRO  was a collaboration of US and Italian
Institutions plus one Moroccan group; see Ref.  \cite{r1} for the list of Authors and Institutions.

\end{document}